\documentclass[acmsmall,screen]{acmart}

\setcopyright{none}             
\renewcommand\footnotetextcopyrightpermission[1]{} 

\author{Lezhi Ma}
\affiliation{
  \institution{Nanjing University}
  \city{Nanjing}
  \country{China}
}
\email{lezhima@hotmail.com}

\author{Shangqing Liu}
\affiliation{
  \institution{Nanjing University}
  \city{Nanjing}
  \country{China}
}
\email{shangqingliu@nju.edu.cn}

\author{Yi Li}
\affiliation{
  \institution{Nanyang Technological University}
  \country{Singapore}
}
\email{yi\_li@ntu.edu.sg}

\author{Qiong Wu}
\affiliation{
  \institution{Nanjing University}
  \city{Nanjing}
  \country{China}
}
\email{qiong937@outlook.com}

\author{Han Wang}
\affiliation{
  \institution{Nanjing University}
  \city{Nanjing}
  \country{China}
}
\email{wh66638@outlook.com}

\author{Jiawan Wang}
\affiliation{
  \institution{Nanjing University}
  \city{Nanjing}
  \country{China}
}
\email{wjwnju@nju.edu.cn}

\author{Lei Bu}
\affiliation{
  \institution{Nanjing University}
  \city{Nanjing}
  \country{China}
}
\email{bulei@nju.edu.cn}

\usepackage{graphicx}
\usepackage{xspace}
\usepackage{tikz}
\usepackage[most]{tcolorbox}
\usepackage{xcolor}
\usepackage{listings}
\usepackage{algpseudocode}
\usepackage[ruled,linesnumbered,vlined]{algorithm2e}
\lstdefinestyle{mystyle}{                
    numbers=left
}
\usepackage{amsmath}
\usepackage{enumerate}
\usepackage{enumitem}
\usepackage{booktabs}
\usepackage{multirow}

\usepackage{amsthm}
\theoremstyle{definition}
\newtheorem{definition}{Definition}
\newtheorem*{definition*}{Definition}
\usepackage{subcaption}

\newcommand{\tool}{\textit{SpecSyn}\xspace}

\begin{document}

\title{\tool: LLM-based Synthesis and Refinement of Formal Specifications for Real-world Program Verification}


\begin{abstract}
Program verification is a formal technique to rigorously ensure the correctness and fault-freeness of software systems. However, constructing comprehensive interprocedural specifications for full verification obligations is time-consuming and labor-intensive, giving rise to automated specification generation approaches. Despite the significant advancements in these approaches enabled by Large Language Models (LLMs), existing LLM-empowered approaches still suffer from significant limitations. Although some approaches can decompose large programs into smaller generation tasks, the strategies are too simplistic to accommodate recursion and other complex control-flow structures. They also typically lack mechanisms to evaluate and guarantee the semantic strength of the generated specifications. The limitations impair their ability to extract precise specifications from complex real-world programs to support the verification of target properties, hindering their applicability to real-world verification tasks. To remedy this gap, we propose \tool, a novel LLM-based specification generation method. \tool first decomposes the input program into individual segments, which are handled respectively by the subsequent iterative specification generation process. Innovatively, we incorporate a specification refinement mechanism into the process, based on semantically non-equivalent program mutations and variant discrimination, to assess and enhance the semantic strength of the generated specifications. Extensive experiments show that \tool maintains high precision over \textbf{90\%} and outstanding recall over \textbf{75\%}, significantly outperforming existing LLM-based approaches. In further evaluations, \tool successfully handles \textbf{1022} out of \textbf{1421} target properties for open-source programs, proving its applicability to real-world program verification tasks.
\end{abstract}


\maketitle              

\section{Introduction}

Program verification is a technique to rigorously ensure the compliance of a given program with certain properties~\cite{d2008survey}, by formalizing the properties into formal program specifications~\cite{leavens2008jml,baudin2008acsl} and proving their correctness with static analysis approaches such as theorem proving~\cite{de2008z3,barrett2011cvc4}. Due to the wide existence of complex control-flow structures (e.g., function calls and loops) within real-world programs, the verification of high-level target properties necessitates interprocedural supporting specifications as a foundation~\cite{wang2025tale}. On account of the intricate and rigorous nature of formal specifications, manually crafting specifications inevitably poses an enormous burden on developers and experts, giving rise to research on automated specification generation techniques. In recent years, the emergence and widespread adoption of Large Language Models (LLMs)~\cite{openai2023gpt,achiam2023gpt,openai2025gpt,guo2025deepseek,grattafiori2024llama} have injected new momentum into automated specification generation approaches~\cite{cao2025informal,ma2024specgen,wang2025tale,wen2024enchanting}, significantly enhancing both the diversity and expressiveness of generated specifications compared to traditional approaches.

Nevertheless, existing LLM-empowered approaches still suffer from limitations in different aspects. Approaches such as AutoSpec~\cite{wen2024enchanting} and SpecGen~\cite{ma2024specgen} treat the input programs holistically and load the entire input program into the context of LLMs for generation. The practice typically triggers the attention dispersion problem~\cite{mudarisov2025limitations} in LLMs and impairs LLM performance as input program length increases, introducing serious scalability issues. Despite the function-level decomposition technique proposed by a concurrent approach, Preguss~\cite{wang2025tale}, the mechanism is primitive and cannot handle circular dependencies between program structures. On the other hand, recent approaches tend to focus primarily on specification correctness, while mechanisms for evaluating and refining the semantic strength of specifications remain largely underexplored. Consequently, the strength of the generated specifications cannot be effectively guaranteed, making them insufficient to support the verification of target properties, limiting the applicability of existing approaches to real-world verification tasks. These limitations pose a fundamental challenge in specification generation: \textit{How to generate precise and useful specifications from sizable, complex real-world programs to effectively support the verification of target properties?}

Confronted with the challenge, we present \tool, a novel LLM-based framework for formal specification generation. Overall, the framework adopts a divide-and-conquer design to enhance its scalability. It first performs static dependency analysis to decompose the input program into a sequence of program segments together with their dependency information. During this process, the specification sketch mechanism is introduced to guide the LLM in analyzing individual program segments and their dependencies, structuring the subsequent generation process. The framework proceeds with iterative specification generation for each program segment, incorporating a novel mechanism for evaluating and refining specification semantic strength based on non-equivalent program mutations and variant discrimination. Guided by the insight that the semantic strength of specifications lies in their ability to distinguish programs from their non-equivalent variants, we measure the semantic strength of specifications by the number of variants refuted and leverage the assessment results to guide the LLM in generating stronger specifications. Eventually, the generated specifications for each segment are synthesized to discharge a complete verification obligation.

To comprehensively evaluate the performance of \tool, we constructed a comprehensive benchmark consisting of 50 self-contained C source files retrieved from various open-source repositories, along with manually constructed ACSL specifications that are strong yet verifiable. Extensive experiments are conducted to compare \tool against a series of state-of-the-art LLM-based approaches. Experimental results show that our method maintains a high level of precision (\textbf{90\%}-\textbf{100\%}), on par with state-of-the-art approaches, while achieving the highest average recall (over \textbf{75\%}) among all baselines, clearly demonstrating the effectiveness of the proposed approach. An ablation study is also conducted to prove the effectiveness of all components within. In further experiments on real-world verification tasks, \tool successfully discharges proof obligations for the most target properties (\textbf{1022} out of \textbf{1421}, compared to 883 of Codex and 510 of Preguss), revealing the practical applicability of our method and the effectiveness of the generated specifications in program verification tasks. In brief, the main contributions of this paper include:
\begin{itemize}[leftmargin=*]
    \item A novel LLM-based framework for specification synthesis to support program verification, incorporating divide-and-conquer style synthesis and a specification strength evaluation and refinement mechanism based on non-equivalent program variant discrimination. 
    \item A new dataset towards specification generation ability evaluation on sizable real-world programs, containing 50 C source files (456.44 LoC on average, with 7 programs exceeding 1000 LoC, compared to programs with fewer than 50 LoC in existing datasets) retrieved from 14 open-source repositories, together with strong yet verifiable ACSL specifications for specification strength assessment, facilitating follow-up research.
    \item A comprehensive evaluation of \tool and state-of-the-art baselines upon the proposed dataset, revealing the effectiveness of \tool in generating verifiable yet semantically strong specifications, along with the applicability of its generated specifications on verification tasks for real-world programs.
\end{itemize}
\section{Background and Motivation}

\subsection{Program Specification Generation and Verification} \label{sec:background}

Program specifications are precise statements that articulate the intended or actual behaviors of a given program~\cite{ma2024specgen}. Notably, formal specifications are formulated in formal languages, such as mathematical notation or Boolean expressions, to define the constraints governing program behavior rigorously. Various formal specification languages have been established for different programming languages. In this work, we primarily focus on the ANSI/ISO C Specification Language (ACSL)~\cite{baudin2008acsl} designed for C programs. Three basic types of specification statements in ACSL are of particular concern: \texttt{requires}, specifying the preconditions on function arguments to guarantee proper function execution; \texttt{ensures}, specifying the postconditions that must hold after function execution; and \texttt{loop invariant}, identifying the properties that consistently hold before executing the loop body.

Benefiting from the syntactic parsability and semantic unambiguity of formal languages, specification verifiers, such as Frama-C/WP~\cite{patrick2025wp} for ACSL, have been developed to validate the correctness of program specifications (i.e., the consistency between specification semantics and the program’s actual behaviors~\cite{ma2024speceval}) through static approaches such as theorem proving and SMT solving~\cite{de2008z3,barrett2011cvc4,conchon2018alt}. Program verification can be achieved by formalizing the desired properties (such as safety constraints and intended functionalities, typically defined by human experts) into specifications and then proving their correctness using the verifiers.
However, successful verification requires more than the target property itself. Real-world programs contain complex control-flow structures that verifiers cannot handle directly, such as loops, function calls, and recursion, necessitating \textit{supporting specifications}, including function contracts for all critical functions relied on and loop invariants for the loops within, to enable compositional verification. Constructing these specifications remains a major challenge, as formal specification languages impose strict syntactic and semantic requirements, making specification crafting both labor-intensive and error-prone, thereby limiting the practical adoption of formal verification.
To alleviate this burden, a long stream of approaches for program specification generation has been proposed, ranging from traditional rule-based methods~\cite{flanagan2001houdini,ernst2007daikon,molina2022fuzzing} and learning–based techniques~\cite{li2017automatic,si2018learning,ryan2019cln2inv} to the more recently emerging approaches~\cite{wen2024enchanting,ma2024specgen,wang2025tale} leveraging Large Language Models (LLMs)~\cite{openai2023gpt,achiam2023gpt,openai2025gpt,liu2024deepseek,guo2025deepseek,grattafiori2024llama}.

\subsection{Motivation} \label{sec:motivation}

\begin{figure*}[!t]
    \centering
    \includegraphics[width=1\linewidth]{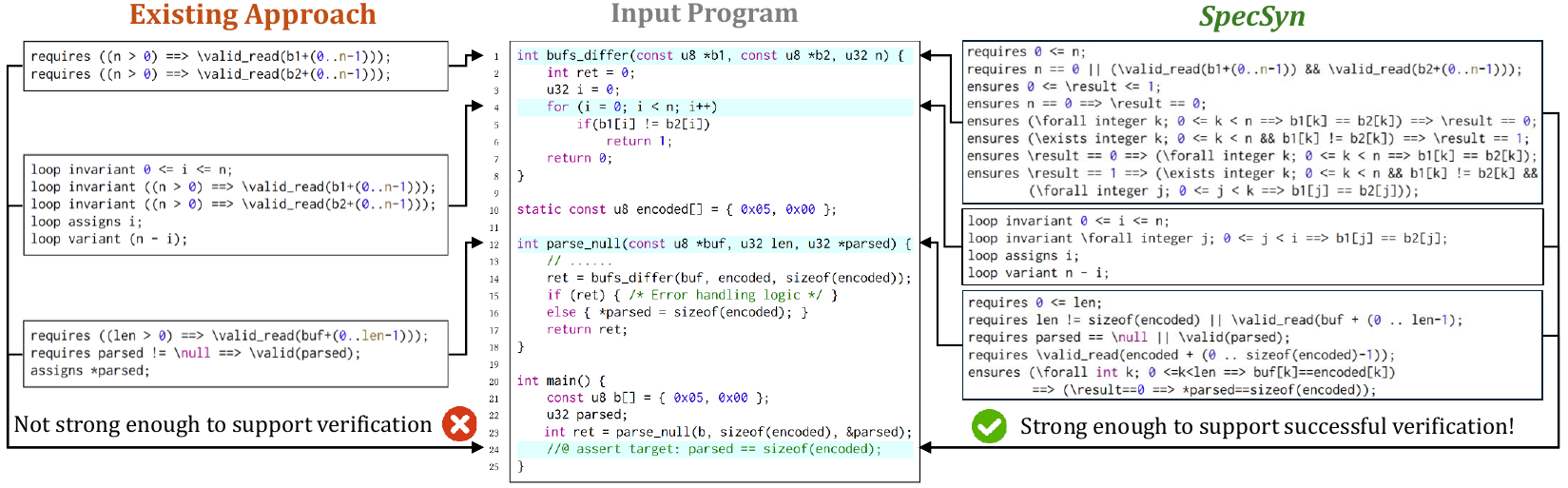}
    \vspace{-1mm}
    \caption{An example program segment from repository \texttt{X509-parser}, the program specifications generated by \tool and a baseline method, and the corresponding verification results of the target property (line 24).}
    \vspace{-4mm}
    \label{fig:motivating_example}
\end{figure*}

Despite the significant advancements in program specification generation approaches brought about by LLMs, existing techniques still suffer from a range of limitations. Some methods, such as AutoSpec~\cite{wen2024enchanting} and SpecGen~\cite{ma2024specgen}, treat the input program holistically, typically presenting the entire program as the target of generation within the context provided to LLMs. This practice has a markedly negative impact on the scalability of the corresponding technique, not only due to the inherent constraint of limited context window sizes on LLMs, but also because of the attention dispersion problem~\cite{mudarisov2025limitations}, where LLMs' ability to distinguish informative tokens declines as the scale of input tokens expands, thereby degrading their performance on specification generation tasks. Despite the naive program decomposition strategy incorporated by certain approaches, they merely process functions independently in sequence, incapable of handling interprocedural control-flow constructs such as recursion. Moreover, existing approaches are primarily driven by the primitive objective of generating specifications that can \textit{pass verification}, while lacking mechanisms to evaluate and enhance specification \textit{strength}, i.e., whether a set of specifications precisely captures the characteristic behaviors of the target program rather than expressing trivial properties that hold for almost any program. Consequently, the strength of the generated specifications is not effectively guaranteed, hindering the application of such specifications and the corresponding generation methods.

One illustrative example of these limitations is presented in Fig.~\ref{fig:motivating_example}, where state-of-the-art approaches~\cite{wen2024enchanting,wang2025tale} showcase unsatisfactory performance in different aspects. Some approaches~\cite{wang2025tale}, driven by the discovery and exclusion of runtime errors, focus solely on memory readability (specified by \texttt{\symbol{92}valid\_read} predicates) issues and fail to generate effective postconditions for the program's essential functionalities, impairing the strength of the generated specifications and causing the verification failure of the target property. Other approaches~\cite{wen2024enchanting} yield only trivial function contracts and loop invariants (e.g., \texttt{0 <= i <= n}) due to the attention dispersion problem, eventually leaving the verification target unprovable. These limitations pose a core challenge on the specification generation task: \textit{how to generate precise and useful specifications from sizable programs, so that the verification of target properties can be effectively supported?}

Confronted with this challenge, we propose \tool, a novel technique featuring the synthesis and refinement of formal specifications for program verification. To alleviate the input scale burden of real-world programs, we introduce \textit{Top-down Task Decomposition} and \textit{Bottom-up Specification Synthesis} to break down the enormous generation task and accomplish it in a divide-and-conquer manner. To strengthen the semantics of the generated specifications, we introduce a specification strength refinement technique based on semantic-non-equivalent program mutation and variant discrimination. Benefited from the introduced designs, \tool succeeded in summarizing comprehensive contracts and invariants for all crucial program points as demonstrated in Fig.~\ref{fig:motivating_example}, laying a solid foundation for the verification of the target property.

\section{Methodology}

\begin{figure}[!t]
    \centering
    \includegraphics[width=0.96\linewidth]{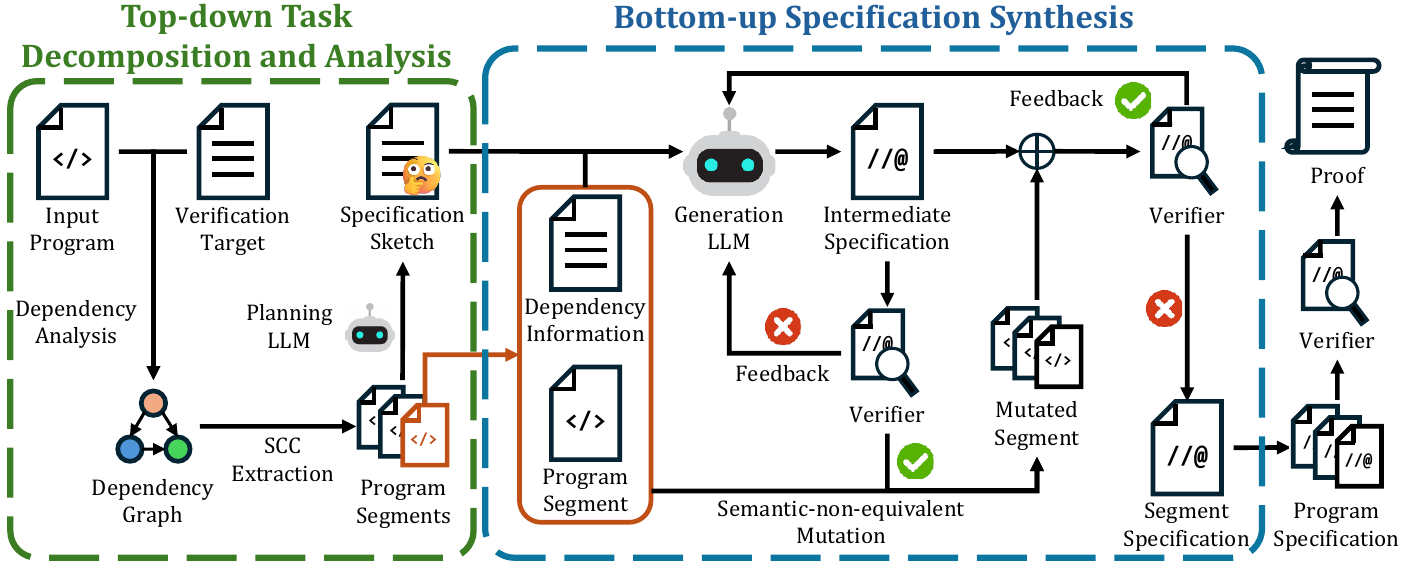}
    \vspace{-2mm}
    \caption{Overview of \tool.}
    \vspace{-4mm}
    \label{fig:overview}
\end{figure}

Fig.~\ref{fig:overview} presents an overview of \tool. Given the source code and verification targets (expressed as ACSL assertions), \tool first performs Top-down Task Decomposition and Analysis (Section~\ref{sec:decomp}), where static analysis constructs a dependency graph whose vertices represent the atomic units of specification generation. It then conducts Bottom-up Specification Synthesis (Section~\ref{sec:synthesis}), iteratively generating specifications for each program segment. After resolving verification errors, a Specification Strength Refinement stage (Section~\ref{sec:refinement}) leverages semantically non-equivalent program mutations to strengthen specification expressiveness and facilitate verification of the target properties. Finally, all generated specifications are integrated to perform complete verification.


\subsection{Top-down Task Decomposition and Analysis} \label{sec:decomp}

\subsubsection{Program Segmentation}

Divide and conquer is a classical paradigm for tackling large-scale problems. When program specifications are to be generated for a large codebase, an intuitive strategy is to decompose it into smaller program segments and process them individually, reducing the amount of input provided to the LLM in each query to alleviate the attention dispersion problem discussed above. Specifically, we begin by extracting all declarations and definitions within the input source code, including function definitions, user-defined type definitions, and free variable declarations, through constructing an Abstract Syntax Tree (AST). The reference relations among these structures (incurred by type references and function calls) are then identified to formulate a dependency graph. To avoid circular dependencies among these structures, we subsequently apply Tarjan’s algorithm~\cite{tarjan1972depth} to identify the strongly connected components (SCCs) of the dependency graph, and treat the corresponding program structures of each SCC as a decomposed segment. Each segment is regarded as a minimal unit for generation and verification. The dependency information between these segments is also summarized according to the dependency graph for further use. It is worth noting that, due to the characteristics of Tarjan’s algorithm, the produced sequence of program segments naturally respects a topological order, i.e., any segment relied on will always appear before the segments that rely on it. Consequently, dependency conflicts are inherently avoided by processing the program segments sequentially according to this order.

\subsubsection{Specification Sketch Generation} 
\begin{figure}[!t]
    \centering
    \includegraphics[width=1\linewidth]{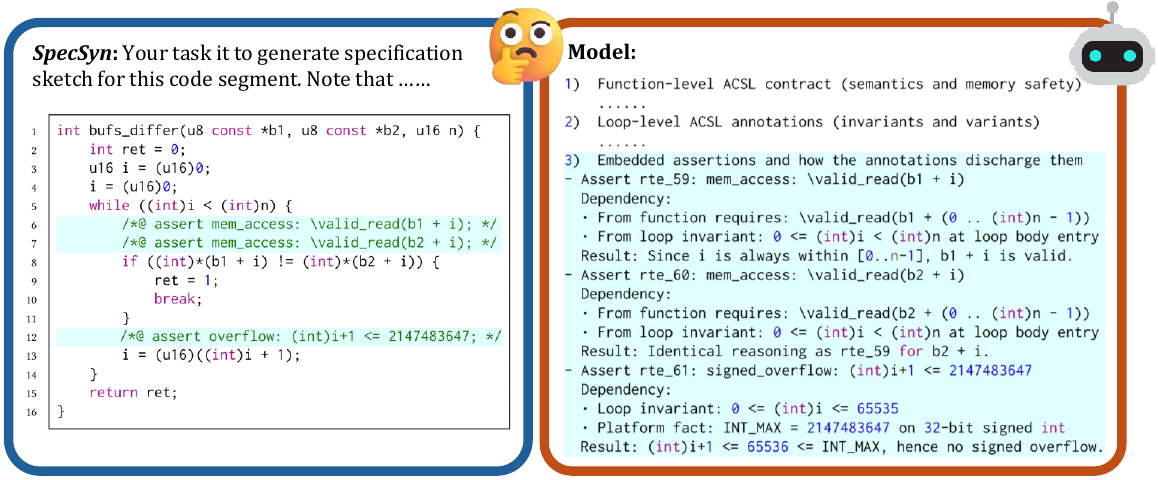}
    \vspace{-5mm}
    \caption{Example specification sketch (as highlighted on the right) for the verification of given target properties (specified as ACSL assertions, as highlighted on the left).}
    \vspace{0mm}
    \label{fig:sketch}
\end{figure}

To better coordinate the specification generation task across the entire program, we introduce the specification sketch mechanism. For each program segment, a planning LLM is instructed to analyze (in natural language) both the semantic content of the segment itself and its dependencies on other segments, and plan the plausible form, syntax structure, and semantic content of the desired specifications. The generated contents are appended to the context in subsequent pipelines as guidance. 
An illustrative example for a specification sketch is presented in Fig.~\ref{fig:sketch}. Towards the three assertions instrumented in the code as verification targets, the LLM performs targeted planning for the preconditions and loop invariants required to support their verification. In the subsequent process, the model will follow this plan to produce the structured specifications for discharging proof obligations towards the targets. This mechanism is inspired by the practice of planning in LLM-based agents, which can enhance the model's comprehension of the assigned task, reduce the risk of hallucinations, and improve the fault tolerance~\cite{huang2024understanding,wang2024survey}. Under the scenario of specification generation and program verification, the mechanism enables the model to reason about the program, verification targets, and corresponding specifications from an interprocedural perspective, rather than focusing solely on the content of an individual program segment, facilitating the preservation of semantic coherence and relevance among the specifications generated for different segments.


\setlength{\textfloatsep}{2mm}

\subsection{Bottom-up Program Specification Synthesis} \label{sec:synthesis}

\begin{algorithm}[!t]
    \scriptsize
    \SetKwFor{Loop}{loop}{}{EndLoop}
    \SetKwInOut{Input}{Input}
    \SetKwInOut{Output}{Output}
    \SetKw{Continue}{continue}
    \SetKw{Return}{return}
    \SetKwIF{If}{ElseIf}{Else}{if}{then}{else if}{else}{endif}
    \Input{Dependency graph $G=\langle V,E \rangle$ of program segments, \\ Hyperparameter $H=\langle n_{refine},n_{repair},t \rangle$, Large Language Model $\mathcal{M}$, verifier $\mathcal{V}$}
    \Output{Set of program specifications $S$}
    \SetKwFunction{BottomUpSynthesis}{BottomUpSynthesis}
    \SetKwFunction{GenSegmentSpecifications}{GenSegmentSpecifications}
    \SetKwProg{Myfunc}{Function}{}{}
    \Myfunc{\BottomUpSynthesis{$G, D, \mathcal{M}, \mathcal{V}$}}{
        $S=\emptyset$ \\
        \For{${v}\in {G.V}$}{
            $S_v = \mathtt{GenSegmentSpecifications}(v,G,D,\mathcal{M}, \mathcal{V})$ \\
            $S=S \cup S_v$
        }
        \Return $S$
    }
    \Myfunc{\GenSegmentSpecifications{$v,G, D, \mathcal{M}, \mathcal{V}$}}{
        $S_v=\emptyset$ \\
        \For{$pos \in \mathtt{ExtractPointsOfInterest}(v.code)$}{
            $S_{pos} = \emptyset$ \\
            $C = \mathtt{AssembleGenerationContext}(v,pos,G)$ \\
            $i=0$ \\
            \Repeat{$n_{refuted}/\mathtt{size}(P_{variant}) \ge H.t$ or $i \ge H.n_{refine}$}{
                $j=0$ \\
                \Repeat{$S_{refuted}$ is $\emptyset$ or $j \ge H.n_{repair}$}{
                    $S_{cand} = \mathcal{M}(C)$ \\
                    $S_{refuted} = \mathcal{V}(S_{cand}, v.code)$ \\
                    $S_{pos} = \mathtt{ClearDuplicate}(S_{pos} \cup (S_{cand} \backslash S_{refuted})$) \\
                    $C = \mathtt{AssembleRepairContext}(C, S_{refuted})$ \\ 
                    $j=j+1$
                }
                $P_{variant} = \mathtt{SemanticNonEquivalentMutate}(v.code)$ \\
                $n_{refuted} = 0$ \\
                $P_{passed} = \emptyset$ \\
                \For{$p_{variant} \in P_{variant}$}{
                    \If{$\mathcal{V}(S_{pos}, p_{variant})$ is not $\emptyset$}{
                        $n_{refuted} = n_{refuted} + 1$
                    }
                    \Else {
                        $P_{passed} = P_{passed} \cup \{p_{variant}\}$
                    }
                }
                $C=\texttt{AssembleRefineContext}(C,P_{passed})$\\
                $i = i + 1$
            }
            $S_v=S_v \cup S_{pos}$
        }
        \Return $S_v$
    }
    \caption{Bottom-up Specification Synthesis}
    \label{alg:bottom_up}
\end{algorithm}

After decomposing the input program into segments, \tool proceeds to generate specifications for each segment and synthesizes the generated specifications to formulate the complete specifications for the entire input program. Algorithm~\ref{alg:bottom_up} presents a detailed illustration of this procedure. The algorithm takes as input a dependency graph $G$ of which the vertices are decomposed program segments and the edges are dependency relations, hyperparameter $H$ specifying the max number of iterations allowed and a threshold for Variant Discrimination Guided Specification Refinement (detailed in section~\ref{sec:refinement}), an LLM $\mathcal{M}$ that outputs a set of specifications according to the given context, and a verifier $\mathcal{V}$ that reads an input program along with specifications and returns a set of unprovable specifications. Generally, as described in function \texttt{BottomUpSynthesis}, the procedure handles each segment (expressed as the nodes $v \in G.V$ within the dependency graph $G$) sequentially according to the aforementioned topological order, and eventually summarizes the specifications of each segment. Function \texttt{GenSegmentSpecifications} implements the process of handling a specific program segment. The function consists of three main components: Points of Interest Identification (line 9 in Alg.~\ref{alg:bottom_up}), Iterative Generation and Repairing (line 14$\sim$21 in Alg.~\ref{alg:bottom_up}), and Variant Discrimination Guided Specification Refinement (line 22$\sim$31 in Alg.~\ref{alg:bottom_up}, detailed in section~\ref{sec:refinement}).

\subsubsection{Points of Interest Identification}

Some existing approaches, e.g., SpecGen~\cite{ma2024specgen}, instruct the LLM to directly regenerate the entire source code with instrumented specifications. This design cannot guarantee preservation of the original program structure due to hallucinations, and specifications at critical locations may be omitted. To address these issues, we first identify all \textit{points of interest} (POIs), i.e., locations where specifications should be inserted (including function entries and loops). The POIs are processed individually in subsequent stages. Notably, dependencies also exist among POIs. For example, proving an outer loop invariant depends on the correctness of inner loop invariants when nested loops occur, and function postconditions rely on the invariants of loops within the function body. To respect these dependencies, we perform a depth-first traversal of the segment AST and use the resulting post-order sequence as the POI processing order.

\subsubsection{Iterative Generation and Repairing}

For each POI, \tool formulates a generation context (line 11 in Alg.~\ref{alg:bottom_up}) consisting of four components: the previously generated specification sketch (Section~\ref{sec:decomp}), the target program segment (with the target POI marked by \texttt{/* >{}>{}>INFILL<{}<{}< */}), task instructions, and the contents of dependent segments. The LLM then generates a set of candidate specifications (line 16). To distinguish correct from incorrect candidates, all candidates are verified against the instrumented program using Frama-C/WP~\cite{patrick2025wp}. Verified candidates are incorporated into the intermediate specification set (line 18), with duplicates removed through lexical analysis. Refuted candidates, together with the reported verification errors, are fed back to the LLM to guide error analysis and repair, producing replacement candidates. This process iterates until all candidates are verified or the maximum number of iterations is reached (line 21).


\subsection{Variant Discrimination Guided Specification Refinement} \label{sec:refinement}

The iterative generation and repairing process typically yields a set of verifiable specification statements. Naturally, one would wish that the generated specifications are semantically related to the target program closely for better support in program verification. Yet, unlike specification correctness, which can be determined relatively easily with the aid of a verifier, the semantic strength of specifications, i.e., \textit{how precise the statements are in terms of reflecting program behaviors}, does not admit a rigorous definition~\cite{polikarpova2013good} and thus cannot be trivially assessed, resulting in the absence of related mechanisms in existing approaches. To bridge this gap, we first propose a specification strength measurement technique based on program mutation and variant discrimination. The idea originates from the concept of \textit{trivial properties} on programs, i.e., properties that either hold for all programs or fail to hold for all programs~\cite{sipser1996introduction}. Similar intuitions can be adapted to program specifications as well: trivial specifications adapt to all programs, while non-trivial specifications hold for some programs but are rejected by others, i.e., they can \textit{distinguish} programs with distinctive semantics. We expect the generated specifications to be more \textit{non-trivial} and \textit{specific} to the target program, so that they are more closely related in terms of semantics. Building on this insight, we propose the concept of \textit{variant discriminative rate} to measure specification strength.
\begin{definition}
    Given a program $p$, a set of \textit{correct} specifications $S$, and a set $M$ of semantic-non-equivalent mutation operators $m:P \rightarrow P$, the variant discriminative rate (VDR) of $S$ towards $M$ is defined as
    \begin{align*}
        r_v(S,p,M)=\frac{\sum\limits_{m \in M}\mathbb{I}(\mathcal{V}(m(p),S)\neq\emptyset)}{|M|}
    \end{align*}
    where $\mathcal{V}:P\times\mathbb{S}\rightarrow\mathbb{S}$ is a verifier that takes as input a program $p \in P$ and a set of specification $S\in\mathbb{S}$ to produce a set of refuted specifications, $\mathbb{I}:\{\mathbf{true},\mathbf{false}\}\rightarrow\{0,1\}$ is the indicator function.
\end{definition}
Based on the definition of VDR, we further formulate the task of \textit{VDR-guided specification refinement} as follows:
\begin{definition}
    Given a program $p \in P$ and a set $M$ of non-equivalent mutation operators, VDR-guided specification refinement is an optimization problem to find a set of specifications $S$ that
    \begin{align*}
        S=\mathop{\arg\min}\limits_{S}\ |M|(1-r_v(S,p,M)),\quad \text{s.t.}\ \mathcal{V}(p,S)=\emptyset
    \end{align*}
    where $\mathcal{V}:P\times\mathbb{S}\rightarrow\mathbb{S}$ is a verifier.
\end{definition}

We employ an iterative process driven by LLMs to address the optimization problem, eventually enabling the evaluation and refinement of specification strength. Generally, for the intermediate result $S_i$ obtained in each iteration, we iteratively optimize the results according to the formula below to yield a new set of specifications $S_{i+1}$:
\begin{align*}
    &S_{i+1} = S_i \cup S_\mathcal{M} \setminus \bigcup_{s \in \mathcal{V}(m(p),S_\mathcal{M})}\{s\} \\
    \mathrm{where} \ & S_\mathcal{M} = \mathcal{M} \left(
        p, S_i, \{m(p)\ |\ m \in M \wedge \mathcal{V}(m(p),S_i)=\emptyset\}
    \right)
\end{align*}
After obtaining a set $S_i$ of verifiable specifications, we apply semantic-non-equivalent mutations on the corresponding code segment to generate a collection of variants (line 22 in Alg.~\ref{alg:bottom_up}). The verifier then attempts to validate the specifications against each variant (line 25$\sim$29 in Alg.~\ref{alg:bottom_up}), with both the number of variants that fail verification (distinguished) and those that pass (undistinguished) recorded. If the proportion of variants that fail verification exceeds a given threshold (line 32 in Alg.~\ref{alg:bottom_up}), the strength of the generated specifications is considered plausible. Otherwise, the undistinguished variants are utilized to construct a new query message (line 30 in Alg.~\ref{alg:bottom_up}), guiding the LLM in generating refined candidates. The message includes a randomly selected undistinguished variant and instructs the LLM to analyze the syntactic and semantic differences between this variant and the original program segment, and to propose new specifications for the properties that explicitly capture these differences. The approach then rolls back to the iterative generation and repairing procedure (section~\ref{sec:synthesis}) to ensure the verifiability of the new candidates.

In terms of the mutation operators, we exploit the 188 mutation operators proposed by Ou et al.~\cite{ou2024mutators}. The operators, originally designed for C compiler fuzzing, are usefully comprehensive, covering a wide range of C language elements, including additions, removals, and modifications on operators, operands, identifiers, statements, control-flow structures, and user-defined type declarations. 35 operators are manually identified as inherently semantically equivalent and are excluded. To further ensure the semantic non-equivalence of the mutations, we adopt Trivial Compiler Equivalence (TCE)~\cite{papadakis2015trivial}, a lightweight technique for fast identification of program equivalence. The principle of TCE is to compile (usually with optimization flags specified) the programs to be assessed and consider them equivalent if the produced binary files are equal. We exploit the GNU C Compiler (GCC)~\cite{GCC} with the \texttt{-O2} optimization flag specified to produce binary files for comparison. Mutated variants detected to be equivalent to the original segment will be excluded from the process.
%


\section{Experimental Setup}

We aim to answer the following research questions:
\begin{itemize}[leftmargin=*]
    \item \textbf{RQ1:} How is the quality of \tool-generated specifications compared to existing approaches?
    \item \textbf{RQ2:} How does each component within \tool contribute to its overall performance?
    \item \textbf{RQ3:} Can \tool-generated specifications solve real-world program verification tasks?
    \item \textbf{RQ4:} How do users perceive the quality of specifications generated by SpecSyn?
\end{itemize}

\vspace{-1mm}
\subsection{Implementation} \label{sec:implementation}

We implement the prototype of \tool in Python for convenient access to LLM interfaces. Three underlying models are adopted, including \texttt{gpt-4-turbo-2024-04-09}, \texttt{gpt-5-2025-08-07}, and DeepSeek-R1.
We exploit the official APIs for access to these models. For all models, temperature is set to 0 whenever allowed to improve performance on reasoning tasks, with other settings left at their defaults. Frama-C/WP~\cite{patrick2025wp}, backed up by Z3~\cite{de2008z3}, is adopted as the verifier. As for the hyperparameter mentioned in Alg.~\ref{alg:bottom_up}, we set $n_{refine}=n_{repair}=5$ and $t=0.75$. All baselines are executed on an Ubuntu 22.04 LTS server with an Intel\textregistered\ Core\texttrademark\ Ultra 9 275HX (2.70 GHz) 24-core processor and 16GB RAM.

\vspace{-1mm}
\subsection{Baselines}

We compare \tool against the following state-of-the-art LLM-based approaches:
\begin{itemize}[leftmargin=*]
    \item \textbf{Preguss}~\cite{wang2025tale}, a specification generation approach featuring the guidance of potential runtime errors (RTEs) extracted by abstract interpretation. 
    \item \textbf{AutoSpec}~\cite{wen2024enchanting}, the first LLM-based specification generation approach to generate and filter candidate specifications. 
    \item \textbf{SpecGen}~\cite{ma2024specgen}, a specification generation approach featuring heuristic mutations performed on the candidate specifications to fix semantic errors. 
    \item \textbf{llama3.1-8b-instruct-fma} and \textbf{qwen2.5-coder-7b-instruct-fma}, models fine-tuned by Cao et al.~\cite{cao2025informal} on data corpora related to program specifications, with performance claimed to be comparable to proprietary models on small-scale specification generation tasks. 
    \item \textbf{Codex}, a state-of-the-art coding agent developed by OpenAI. When executing the tasks, Codex is allowed to read/write source files and invoke Frama-C, without permission to access the internet. Note that Codex is only evaluated with GPT-5 as the backbone LLM, since the agentic decision-making ability of DeepSeek-R1 and GPT-4 is insufficient to ensure robust execution.
\end{itemize}

\vspace{-1mm}
\subsection{Benchmark Construction} \label{sec:benchmark}


\subsubsection{RQ1\&2} Currently, few specification synthesis benchmarks targeting real-world programs have been proposed. Despite the existence of certain datasets~\cite{githubGitHubManavpatnaikframacproblems,githubGitHubFraunhoferfokusacslbyexample} for toy examples, the programs contained are usually small in scale (typically less than 50 LoC per program), and the verifiability of the specifications provided is not strictly guaranteed, making them inadequate for reflecting the performance of baselines on real-world programs. To support a comprehensive evaluation, we extracted a series of self-contained C source files from the following real-world repositories:
\begin{itemize}[leftmargin=*]
    \item \textbf{UAV-Quadcopter}~\cite{githubGitHubSHRCUAVQuadcopter}, an UAV controlling system based on Arduino. 1 file concerning quaternion operations is retrieved. 
    \item \textbf{voronoi}~\cite{githubGitHubJCashvoronoi}, a Voronoi diagram generation tool using Fortune's algorithm~\cite{fortune1986sweepline}. 2 files concerning basic graph structures and relevant operations are retrieved from the repository. 
    \item \textbf{Hypatia}~\cite{hypatia}, a low earth orbit satellite network simulation framework. 1 file on vector and matrix operations is retrieved. 
    \item \textbf{sokol}~\cite{githubGitHubFlooohsokol} and \textbf{CGL}~\cite{githubGitHubJaysmito101cgl}, development kits targeting WebAssembly. 3 files concerning color and shape processing are retrieved.
    \item \textbf{BlueShiftEngine}~\cite{githubGitHubPolygonTekBlueshiftEngine}, \textbf{Maratis}~\cite{maratis3d}, and \textbf{Punity}~\cite{githubGitHubMartincohenPunity}, rendering tools for demo construction and game development. 7 files concerning fast algorithms and data structures for camera control and image processing are retrieved. 
    \item \textbf{ccan}~\cite{githubGitHubRustyrussellccan}, \textbf{clib}~\cite{githubGitHubClibsclib}, \textbf{zpl}~\cite{githubGitHubZplczpl}, and \textbf{stb}~\cite{githubGitHubNothingsstb}, third-party public domain libraries for utilities covering multiple areas. We retrieved a total of 22 files from these repositories that serve a wide spectrum of purposes, including basic data structures, string operations, timestamp management, and parser components.
\end{itemize}
A group of 3 Ph.D. students, all with at least 4 years of C programming experience and trained on ACSL specifications and Frama-C/WP usage, participated in the construction under the instructions of a strict, predefined protocol as follows. 

\noindent\textbf{Repository selection.} We define the target population as open-source C projects that implement practical software functionalities. Candidate repositories are retrieved from GitHub according to predefined criteria (pure C language, active in the past 1 year, suitably licensed, over 1000 LoC). Repositories that cannot be parsed by Frama-C are automatically detected and excluded. The remaining ones are manually inspected to exclude deprecated repositories, educational projects, forked projects, and toy examples/benchmarks.

\noindent\textbf{Function selection.} Appropriate functions should expose well-defined, deterministic functional behavior and be amenable to ACSL verification. We manually excluded trivial wrappers, debugging utilities, logging functions, macro expansions, and functions whose semantics depend on operating-system interactions. Repositories in which more than 80\% of the code cannot be adequately specified using ACSL are also excluded.

\noindent\textbf{Specification crafting.} For each valid function, we construct ground-truth specifications through an iterative workflow. Initial specifications are drafted based on the function implementation and are subsequently refined through multiple rounds of verification using Frama-C, where repeated revisions are conducted to eliminate inconsistent statements and missing behaviors. All specifications crafted undergo cross-validation involving all 3 students to ensure their semantic strength. If any participant raises concern that certain program behaviors may not be covered, the specifications must be revised for better relevance to target code behaviors and undergo the review process again, until all participants reach agreement on the sufficiency of specification strength. 


Apart from these programs, we also utilize the following established specifications from existing works~\cite{wen2024enchanting,Nilizadeh2021AprFormalMethods}: \textbf{X509-parser}~\cite{githubGitHubANSSIFRx509parser}, an RTE-free parser for X.509 certificates~\cite{rfc5280}, and \textbf{Java-JML}~\cite{Nilizadeh2021AprFormalMethods}, a set of Java programs with verifiable specifications (translated to C and ACSL). A total of 50 programs were eventually collected. 


\subsubsection{RQ3} Further evaluation towards software verification tasks necessitates programs with corresponding verification targets. We selected 3 files from \textbf{voronoi}~\cite{githubGitHubJCashvoronoi}, \textbf{Hypatia}~\cite{hypatia}, and \textbf{sokol}~\cite{githubGitHubFlooohsokol} since they are equipped with relevant documentation and test suites to construct the verification targets needed. 3 Ph.D. students are employed to manually translate the contents into ACSL assertions as targets, following the same iterative revision procedure described in section~\ref{sec:benchmark}. The targets are double-validated by Frama-C/WP to guarantee their correctness and verifiability. Apart from these repositories, we also utilize source files from \textbf{X509-parser}~\cite{ebalard2019journey}, \textbf{Contiki}~\cite{peyrard2018towards}, and \textbf{Atomthreads}~\cite{beyer2025improvements}, with RTE-related assertions retrieved from the Preguss~\cite{wang2025tale} artifact as targets. Additionally, we adopt the 213 programs in \textbf{SV-COMP Benchmark} (Reachability category), summarized by Cao et al.~\cite {cao2025informal}, in which the original assertions are already translated into ACSL statements without modification.

\vspace{-1mm}
\subsection{Metrics}

To comprehensively evaluate the quality of the generated specifications in terms of both \textit{correctness} and \textit{strength}, we follow the settings of prior works~\cite{molina2021evospex,molina2022fuzzing} and adopt the following two metrics for performance evaluation.

\noindent\textbf{Precision}, defined by the percentage of \textit{correct} specifications over all generated specifications, measuring the \textit{correctness} of these specifications. To determine whether a specification is \textit{correct}, we extract each specification from the generated results and insert it into the same spot within the source code instrumented with ground-truth specifications, formulating a verification harness. The inserted specification is considered correct if the verifier accepts the verification harness.

\noindent\textbf{Recall}, defined by the percentage of ground-truth specifications \textit{covered by the generated specifications} over all ground-truth specifications, measuring the \textit{strength} of the generated specifications. Since the ground-truth specifications are manually crafted to guarantee a precise reflection of program behaviors, they are considered \textit{strong} enough to serve as a standard for specification strength. To determine whether a ground-truth specification $\hat{s}$ is covered, we utilize the set $S$ of generated specifications on the same program point to construct a composite specification $\tilde{s} = (\hat{s} \leftrightarrow \bigwedge_{s \in S} s)$ and check the correctness of $\tilde{s}$ with the verifier. $\hat{s}$ is considered covered if $\tilde{s}$ is verifiable.

\section{Experimental Results}

\subsection{RQ1: Overall Performance of Specification Generation} \label{sec:RQ1}

\begin{table*}[!t] 
\centering
\caption{Precision (P.) and Recall (R.) of the specifications generated by each baseline (in percentages).}
\vspace{-2mm}
\resizebox{\textwidth}{!}{
\begin{tabular}{ccc|cccccccccccccc|c}
\toprule
\multicolumn{3}{c|}{Repository} & X509 & Java-JML & UAV & voronoi & hypatia & sokol & cgl & BSE & Maratis & Punity & ccan & clib & zpl & stb & \multirow{2}{*}{Avg.} \\ \cmidrule{1-17}
\multicolumn{3}{c|}{Average LoC per File} & 95 & 207 & 1741 & 348 & 1323 & 553 & 350 & 358 & 732 & 550 & 632 & 551 & 524 & 299 &  \\ \midrule
\multicolumn{1}{c|}{\multirow{6}{*}{\tool}} & \multicolumn{1}{c|}{\multirow{2}{*}{GPT-5}} & P. & 95.66 & 97.80 & 97.57 & 99.39 & 97.24 & 93.42 & \textbf{100.00} & 98.18 & \textbf{99.13} & \textbf{98.77} & \textbf{99.24} & 90.94 & 86.88 & \textbf{99.31} & \textbf{96.68} \\
\multicolumn{1}{c|}{} & \multicolumn{1}{c|}{} & R. & \textbf{85.30} & 82.27 & 41.96 & \textbf{86.83} & \textbf{95.67} & 76.77 & 95.83 & 73.09 & \textbf{84.53} & 47.37 & \textbf{80.58} & \textbf{77.43} & \textbf{73.48} & \textbf{61.62} & \textbf{75.91} \\ \cmidrule{2-18} 
\multicolumn{1}{c|}{} & \multicolumn{1}{c|}{\multirow{2}{*}{GPT-4}} & P. & \textbf{98.61} & \textbf{98.37} & 95.69 & 98.43 & 73.81 & 89.29 & 100.00 & 93.91 & 97.15 & 95.95 & 98.89 & 95.49 & 91.08 & 96.61 & 94.52 \\
\multicolumn{1}{c|}{} & \multicolumn{1}{c|}{} & R. & 64.17 & 76.51 & 40.79 & 86.83 & 81.10 & 79.21 & \textbf{100.00} & \textbf{77.91} & 56.86 & \textbf{80.70} & 66.19 & 61.99 & 66.29 & 56.89 & 71.10 \\ \cmidrule{2-18} 
\multicolumn{1}{c|}{} & \multicolumn{1}{c|}{\multirow{2}{*}{DS-R1}} & P. & 96.11 & 96.07 & 98.87 & 93.34 & 89.74 & 99.13 & 100.00 & 94.67 & 98.80 & 91.88 & 98.76 & 93.12 & \textbf{92.79} & 97.86 & 95.80 \\
\multicolumn{1}{c|}{} & \multicolumn{1}{c|}{} & R. & 36.95 & 44.61 & \textbf{49.42} & 83.83 & 92.91 & 33.65 & 100.00 & 56.80 & 58.48 & 57.89 & 61.49 & 58.22 & 63.98 & 58.57 & 61.20 \\ \midrule
\multicolumn{1}{c|}{\multirow{4}{*}{AutoSpec}} & \multicolumn{1}{c|}{\multirow{2}{*}{GPT-5}} & P. & 33.33 & 84.72 & \textbf{100.00} & 80.00 & 95.24 & 81.25 & 42.86 & 85.40 & 16.67 & 90.00 & 66.67 & 79.39 & 81.67 & 0.00 & 66.94 \\
\multicolumn{1}{c|}{} & \multicolumn{1}{c|}{} & R. & 9.52 & 28.21 & 15.85 & 31.62 & 1.57 & 6.72 & 66.67 & 10.31 & 0.84 & 14.04 & 11.98 & 21.56 & 11.87 & 0.00 & 16.48 \\ \cmidrule{2-18} 
\multicolumn{1}{c|}{} & \multicolumn{1}{c|}{\multirow{2}{*}{GPT-4}} & P. & 74.21 & 78.71 & 50.00 & 69.55 & 94.59 & 37.50 & 40.85 & 75.43 & 80.00 & 45.77 & 81.11 & 71.60 & 47.94 & 81.03 & 66.31 \\
\multicolumn{1}{c|}{} & \multicolumn{1}{c|}{} & R. & 33.86 & 26.31 & 15.85 & 34.62 & 28.35 & 1.32 & 29.17 & 26.60 & 21.00 & 5.26 & 41.47 & 16.54 & 10.55 & 18.75 & 22.12 \\ \midrule
\multicolumn{1}{c|}{\multirow{4}{*}{SpecGen}} & \multicolumn{1}{c|}{\multirow{2}{*}{GPT-5}} & P. & 75.97 & 76.50 & 53.85 & \textbf{100.00} & 69.67 & 91.89 & 100.00 & 90.58 & 97.67 & 93.85 & 69.84 & 78.63 & 80.05 & 88.71 & 83.37 \\
\multicolumn{1}{c|}{} & \multicolumn{1}{c|}{} & R. & 43.46 & 68.20 & 15.38 & 77.12 & 51.97 & \textbf{81.80} & 100.00 & 79.73 & 58.06 & 68.42 & 44.19 & 50.45 & 61.24 & 43.19 & 60.23 \\ \cmidrule{2-18} 
\multicolumn{1}{c|}{} & \multicolumn{1}{c|}{\multirow{2}{*}{GPT-4}} & P. & 92.21 & 75.84 & 50.00 & 90.68 & 98.51 & 0.00 & 100.00 & 12.50 & 89.81 & 98.25 & 91.07 & 80.63 & 66.52 & 70.73 & 72.62 \\
\multicolumn{1}{c|}{} & \multicolumn{1}{c|}{} & R. & 32.06 & 42.94 & 14.69 & 84.29 & 61.42 & 0.00 & 100.00 & 8.70 & 48.28 & 82.46 & 45.26 & 35.76 & 38.77 & 34.54 & 44.94 \\ \midrule
\multicolumn{1}{c|}{\multirow{4}{*}{Preguss}} & \multicolumn{1}{c|}{\multirow{2}{*}{GPT-5}} & P. & 66.67 & 95.78 & 99.54 & 100.00 & 99.13 & 99.00 & 90.00 & \textbf{100.00} & 97.38 & 97.96 & 81.98 & \textbf{96.56} & 89.73 & 98.15 & 93.70 \\
\multicolumn{1}{c|}{} & \multicolumn{1}{c|}{} & R. & 10.27 & 29.15 & 35.43 & 36.11 & 53.15 & 13.85 & 33.33 & 21.17 & 32.76 & 24.56 & 36.42 & 38.68 & 36.44 & 28.69 & 30.72 \\ \cmidrule{2-18} 
\multicolumn{1}{c|}{} & \multicolumn{1}{c|}{\multirow{2}{*}{GPT-4}} & P. & 66.67 & 98.09 & 99.28 & 100.00 & \textbf{100.00} & \textbf{100.00} & 100.00 & 100.00 & 97.53 & 98.25 & 83.33 & 96.37 & 86.56 & 92.82 & 94.21 \\
\multicolumn{1}{c|}{} & \multicolumn{1}{c|}{} & R. & 15.61 & 24.85 & 35.66 & 23.73 & 34.25 & 15.16 & 37.50 & 32.14 & 27.14 & 22.81 & 24.76 & 28.93 & 27.93 & 25.56 & 26.86 \\ \midrule
\multicolumn{1}{c|}{\multirow{2}{*}{Codex}} & \multicolumn{1}{c|}{\multirow{2}{*}{GPT-5}} & P. & 85.61 & 98.31 & 56.30 & 93.05 & 84.28 & 95.84 & 100.00 & 99.45 & 85.20 & 86.54 & 95.54 & 90.79 & 72.20 & 87.50 & 87.90 \\
\multicolumn{1}{c|}{} & \multicolumn{1}{c|}{} & R. & 58.16 & \textbf{92.90} & 25.41 & 85.60 & 67.70 & 64.70 & 83.33 & 69.78 & 72.35 & 73.68 & 73.68 & 86.59 & 54.72 & 44.07 & 68.05 \\ \midrule
\multicolumn{2}{c|}{\multirow{2}{*}{llama3.1-fma}} & P. & 25.24 & 26.75 & 0.00 & 0.00 & 0.00 & 21.55 & 0.00 & 0.00 & 0.00 & 0.00 & 0.00 & 0.00 & 0.00 & 0.00 & 5.25 \\
\multicolumn{2}{c|}{} & R. & 0.00 & 11.97 & 0.00 & 0.00 & 0.00 & 10.81 & 0.00 & 0.00 & 0.00 & 0.00 & 0.00 & 0.00 & 0.00 & 0.00 & 1.63 \\ \midrule
\multicolumn{2}{c|}{\multirow{2}{*}{qwen2.5-fma}} & P. & 0.00 & 7.74 & 0.00 & 0.00 & 0.00 & 0.00 & 0.00 & 0.00 & 0.00 & 0.00 & 0.00 & 0.00 & 0.00 & 0.00 & 0.55 \\
\multicolumn{2}{c|}{} & R. & 0.00 & 2.73 & 0.00 & 0.00 & 0.00 & 0.00 & 0.00 & 0.00 & 0.00 & 0.00 & 0.00 & 0.00 & 0.00 & 0.00 & 0.20 \\ \bottomrule
\end{tabular}
}
\label{tab:RQ1}
\end{table*}

Table~\ref{tab:RQ1} shows the quality (correctness measured by precision, and strength measured by recall) of the specifications generated by each baseline method on the constructed dataset.

\subsubsection{Specification Correctness}
Overall, \tool incorporating GPT-5 yields the highest average precision (96.68\%) among all baseline approaches. The following baseline approach is Preguss, with both the GPT-5 and GPT-4 versions yielding a precision above 93\%. Both approaches are equipped with effective mechanisms to filter out incorrect specifications according to the feedback of the verifier, guaranteeing the correctness of the generated specifications. Consequently, the difference in precision between the two methods is not substantial. In contrast to these methods, the remaining approaches suffer from a substantial degradation in precision. SpecGen achieves 83.37\% precision, while AutoSpec only yields 66.94\%. As claimed in section~\ref{sec:motivation}, these two approaches treat the input program holistically, loading the entire program into LLMs' context, which hinders the performance of LLMs and the quality of generated specifications. Specifically, for programs over 500 LoC, AutoSpec is only able to generate specifications for a handful of functions, and the output content of the incorporated LLM typically suffers from serious syntax errors. The remaining two fine-tuned models, lama3.1-8b-instruct-fma and qwen2.5-coder-7b-instruct-fma, can only generate limited meaningful content on the input programs and thus yield marginal precision. The two models are fine-tuned on relatively small-scale (7B$\sim$8B parameters) base models, which inherently struggle to handle sizable input.

\subsubsection{Specification Strength}
Generally, \tool incorporating GPT-5 yields the highest average recall (75.91\%) on all programs adopted in the experiments, significantly outperforming all other baselines involved. Other versions of \tool incorporating GPT-4 and Deepseek-R1 also demonstrate considerable performance, achieving 71.10\% and 61.20\% recall, respectively. Notably, for 12 repositories among the 14 involved in the experiments, the highest average recall is observed on the different versions of \tool. The result indicates the remarkable semantic strength of the specifications by \tool, benefiting from the innovative mechanism for specification refinement based on program mutation and variant discrimination. Apart from \tool, the best-performing approach is Codex incorporating GPT-5. In comparison, Preguss can only achieve a relatively limited recall of around 30\%. Preguss places excessive emphasis on RTE guidance, which is closely tied to preconditions but has only an indirect influence on postconditions and loop invariants, and thus fails to provide strong, effective guidance for these two types of specifications, adversely affecting their quantity and quality. AutoSpec is constrained by the LLM’s attention dispersion problem and consequently yields only around 20\% recall. The recall of the two fine-tuned models remains exceptionally low due to the aforementioned scalability problem of small-scale models.

\subsubsection{Consistency Across Repositories}
\tool demonstrates consistent specification synthesis performance across diverse repositories, achieving high precision and recall on the majority of the evaluated projects. In particular, GPT-5 achieves a precision exceeding 90\% on all 14 repositories, while its recall exceeds 70\% on 11 of them, indicating its ability to generate both correct and comprehensive specifications across a wide range of codebases with substantially different code sizes and program characteristics. In contrast, existing baselines exhibit more pronounced performance variations across repositories. For instance, AutoSpec and Preguss achieve particularly low recall on several projects despite maintaining high precision, while SpecGen exhibits substantial fluctuations in both metrics across repositories. These results highlight the robustness of \tool across diverse software projects and its ability to consistently synthesize high-quality specifications.

\subsubsection{Underlying Model Impact}
The choice of underlying LLM has a notable impact on specification synthesis quality. Among the three evaluated models, GPT-5 achieves the highest average precision and recall of 96.68\% and 75.91\%, respectively, followed by GPT-4 (94.52\% and 71.10\%) and DeepSeek-R1 (95.80\% and 61.20\%). While the models exhibit relatively comparable precision, their recall differs more substantially, suggesting that the underlying LLM has a greater influence on specification completeness than correctness within \tool. Nevertheless, model performance varies across repositories: for instance, DeepSeek-R1 achieves a higher recall than GPT-5 on UAV (49.42\% vs. 41.96\%), whereas GPT-5 performs better on Java-JML (82.27\% vs. 44.61\%). These findings suggest that although stronger underlying models can improve overall specification synthesis quality, their effectiveness remains dependent on the characteristics of the target codebase.

\vspace{-1mm}
\begin{tcolorbox}[left=1mm,right=1mm]
    \vspace{-1mm}
    \textbf{Answer to RQ1:} \tool substantially outperforms all state-of-the-art baselines in both specification correctness (96.68\% precision) and strength (75.91\% recall), while being able to maintain performance consistency across the majority of repositories. Its superior performance demonstrates the effectiveness of combining a divide-and-conquer strategy with iterative refinement for scalable and high-quality specification synthesis.
    \vspace{-1mm}
\end{tcolorbox}
\vspace{-1mm}

\subsection{RQ2: Ablation Study} \label{sec:RQ2}

\begin{table*}[!t] 
\centering
\caption{Precision (P.) and Recall (R.) of \tool with main components disabled, respectively.}
\vspace{-2mm}
\resizebox{\textwidth}{!}{
\begin{tabular}{@{}cc|cccccccccccccc|c@{}}
\toprule
\multicolumn{2}{c|}{Repository} & X509 & Java-JML & UAV & voronoi & hypatia & sokol & cgl & BSE & Maratis & Punity & ccan & clib & zpl & stb & \multirow{2}{*}{Avg.} \\ \cmidrule(r){1-16}
\multicolumn{2}{c|}{Average LoC} & 95 & 207 & 1741 & 348 & 1323 & 553 & 350 & 358 & 732 & 550 & 632 & 551 & 524 & 299 &  \\ \midrule
\multicolumn{1}{c|}{\multirow{2}{*}{\tool}} & P. & 95.66 & \textbf{97.80} & \textbf{97.57} & \textbf{99.39} & \textbf{97.24} & \textbf{93.42} & \textbf{100.00} & \textbf{98.18} & \textbf{99.13} & \textbf{98.77} & \textbf{99.24} & \textbf{90.94} & \textbf{86.88} & \textbf{99.31} & \textbf{96.68} \\
\multicolumn{1}{c|}{} & R. & \textbf{85.30} & \textbf{82.27} & \textbf{41.96} & 86.83 & \textbf{95.67} & \textbf{76.77} & \textbf{95.83} & \textbf{73.09} & \textbf{84.53} & 47.37 & \textbf{80.58} & \textbf{77.43} & \textbf{73.48} & \textbf{61.62} & \textbf{75.91} \\ \midrule
\multicolumn{1}{c|}{\multirow{2}{*}{\begin{tabular}[c]{@{}c@{}}w/o\\ decomposition\end{tabular}}} & P. & \textbf{98.58} & 97.64 & 66.56 & 98.52 & 96.54 & 88.47 & 99.40 & 25.00 & 47.76 & 98.42 & 98.28 & 52.38 & 15.97 & 98.57 & 77.29 \\
\multicolumn{1}{c|}{} & R. & 84.23 & 80.82 & 21.21 & \textbf{88.98} & 77.17 & 71.91 & 91.67 & 23.19 & 31.81 & 52.63 & 79.15 & 39.46 & 19.45 & 80.93 & 60.19 \\ \midrule
\multicolumn{1}{c|}{\multirow{2}{*}{\begin{tabular}[c]{@{}c@{}}w/o\\ refinement\end{tabular}}} & P. & 84.09 & 80.90 & 82.86 & 75.40 & 85.99 & 83.75 & 96.49 & 88.96 & 92.13 & 92.21 & 87.26 & 79.15 & 81.28 & 83.56 & 85.29 \\
\multicolumn{1}{c|}{} & R. & 73.57 & 69.10 & 18.65 & 84.29 & 56.30 & 68.76 & 91.67 & 76.00 & 77.07 & 52.63 & 59.47 & 61.34 & 61.32 & 35.26 & 63.25 \\ \midrule
\multicolumn{1}{c|}{\multirow{2}{*}{\begin{tabular}[c]{@{}c@{}}w/o\\ sketch\end{tabular}}} & P. & 98.08 & 97.64 & 99.51 & 99.31 & 96.54 & 90.37 & 98.80 & 95.82 & 98.01 & 98.42 & 95.07 & 80.10 & 79.27 & 97.86 & 94.63 \\
\multicolumn{1}{c|}{} & R. & 84.23 & 81.04 & 39.63 & 85.14 & 77.95 & 75.89 & 91.67 & 58.20 & 83.01 & \textbf{57.89} & 70.99 & 56.84 & 65.04 & 60.42 & 70.57 \\ \bottomrule
\end{tabular}
}
\label{tab:RQ2}
\end{table*}

We further conduct an ablation study to evaluate the effectiveness of the main components within \tool. The results are shown in Table~\ref{tab:RQ2}, where w/o \{*\} denotes the variant of \tool with the corresponding component disabled. All variants incorporate GPT-5 as the underlying LLM.

\subsubsection{Overall Component Contributions}
The ablation results demonstrate that all three components contribute to the effectiveness of \tool, albeit to different extents. Removing decomposition leads to the most substantial performance degradation, reducing average precision and recall by 19.39 and 15.72 percentage points, respectively. This highlights the importance of decomposing specification synthesis into manageable, interdependent tasks. Removing refinement also results in considerable drops of 11.39 and 12.66 percentage points in precision and recall, respectively, underscoring the role of iterative verification-guided repair and critique-guided refinement in improving specification quality. In comparison, removing sketch yields relatively modest average declines of 2.05 and 5.34 percentage points, suggesting that sketch primarily provides complementary guidance for specification coverage rather than being the primary driver of correctness.

\subsubsection{Repository-level Impact}
The contributions of these components vary across repositories, reflecting differences in program structures and specification requirements. Decomposition is particularly beneficial for BlueShiftEngine and zpl, where its removal causes substantial declines in both precision and recall, indicating its importance in managing structural and specification dependencies. Refinement, in contrast, delivers more consistent benefits across repositories, improving precision in all 14 cases and yielding particularly notable recall gains on UAV and hypatia. Although sketch has a smaller overall impact, it substantially improves recall on repositories such as clib, hypatia, and BlueShiftEngine, demonstrating its value in guiding specification coverage for certain programs. Meanwhile, occasional improvements in individual metrics after ablation suggest that the benefits of each component are influenced by repository-specific characteristics, rather than being uniformly realized across all programs.

\vspace{-1mm}
\begin{tcolorbox}[left=1mm,right=1mm]
    \vspace{-1mm}
    \textbf{Answer to RQ2:} The ablation study confirms that all three components contribute to \tool's specification synthesis capability. Decomposition has the largest impact on overall performance, while refinement consistently improves specification correctness and strength. Sketch provides complementary benefits, particularly in improving specification coverage on certain repositories. These results demonstrate that the three components play distinct yet complementary roles in achieving high-quality specification synthesis.
    \vspace{-1mm}
\end{tcolorbox}
\vspace{-1mm}


\subsection{RQ3: Real-world Software Verification Tasks} \label{sec:RQ3}

To evaluate the practical effectiveness of the generated specifications in real-world verification tasks, we assess whether the specifications produced by each baseline can successfully discharge verification targets across seven benchmarks. As shown in Table~\ref{tab:RQ3}, \tool successfully discharges 1,022 out of 1,421 verification targets, compared with 883 for Codex, 510 for Preguss, and 259 for AutoSpec. This demonstrates the effectiveness of \tool across diverse verification tasks, with consistent gains over Codex, another agentic approach, in the overall number of successfully verified targets. Specifically, the programs fall into three categories.

\begin{table}[!t] 
\centering
\caption{Number of proved target properties by each baseline method.}
\vspace{-3mm}
\resizebox{0.83\linewidth}{!}{
\begin{tabular}{c|ccc|ccc|c|c}
\toprule
Repository & voronoi & Hypatia & sokol & X509-parser & Contiki & Atomthreads & SVCOMP & Total \\ \midrule
Avg. LoC per File & 424 & 1323 & 821 & 1199 & 544 & 1451 & 27.79 & - \\
Number of Targets & 172 & 195 & 241 & 142 & 163 & 239 & 269 & 1421 \\ \midrule
\tool & \textbf{153} & \textbf{187} & \textbf{220} & 120 & \textbf{105} & 99 & \textbf{138} & \textbf{1022} \\
Preguss & 39 & 15 & 31 & \textbf{125} & 103 & \textbf{109} & 88 & 510 \\
AutoSpec & 19 & 22 & 13 & 81 & 89 & 7 & 28 & 259 \\
SpecGen & 110 & 71 & 93 & 41 & 43 & 27 & 90 & 475 \\
Codex & 139 & 165 & 171 & 101 & 96 & 90 & 121 & 883 \\
llama3.1-fma & 9 & 0 & 14 & 0 & 0 & 0 & 6 & 29 \\ \bottomrule
\end{tabular}
}
\label{tab:RQ3}
\end{table}


{\noindent\bf Voronoi, Hypatia, and Sokol.}
Constructed according to the software documentation and test suites of the original repositories, the targets within these programs mainly focus on functional correctness and encompass a wide spectrum of properties, including checks on variable values, boundary conditions, pointer reference validity, and universal/existential property assertions. Proof obligations for these targets necessitate a relatively comprehensive summarization of program functionalities and behaviors. On these programs, our method demonstrates outstanding performance, successfully discharging obligations for around 90\% of verification targets for each program. This benefits from the specification refinement mechanism and the strong specifications produced, which can precisely capture the essential behaviors and functionalities of the input programs, formulating a solid foundation for verification of target properties. In contrast, Preguss can discharge only 10\%$\sim$25\% of the proof obligations, as its generation process focuses solely on RTEs, causing insufficient generality in the generated specifications. Despite the relatively competitive recall achieved by SpecGen, its practice of directly instructing the LLM to regenerate the entire source code occasionally causes the model to hallucinate modifications to the original program structure, leaving some verification targets invalid. Suffering from the scalability problem introduced in section~\ref{sec:motivation}, AutoSpec and \texttt{llama3.1-8b-instruct-fma} can hardly generate meaningful specifications, struggling to support the verification.


{\noindent\bf X509-parser, Contiki, and Atomthreads.}
These programs are retrieved from the artifact of Preguss~\cite {wang2025tale}, an RTE-oriented approach, and the corresponding targets are assertions that exclude RTEs from execution. Preguss demonstrates considerable capability in handling these targets, generating specifications that successfully support the verification of 337 out of 544 targets by meticulously handling the inference of appropriate preconditions. In comparison, without applying any specialized mechanism towards RTE, \tool still achieves performance comparable to Preguss, successfully yielding proof obligations for 324 targets. This benefits from the specification sketch introduced in section~\ref{sec:decomp}, where the LLM identifies the assertions instrumented in the input code and plans their verification in advance.


{\noindent\bf SV-COMP.}
Most programs in the SV-COMP benchmark are relatively small in scale (27.79 LoC on average for the programs adopted for experiments) and rarely involve function calls, with their verification primarily relying on appropriate loop invariants rather than interprocedural specifications. On these programs, \tool still showcases leading performance among all baseline methods, successfully discharging verification obligations for 138 out of 269 targets. In contrast, the capability of Preguss and SpecGen in inferring loop invariants can only help verify a limited subset of targets. Despite the significant reduction in the size of the input programs, the specifications generated by AutoSpec and \texttt{llama3.1-8b-instruct-fma} remain largely insufficient to support the verification of the target properties, since their design does not take proof obligation production into consideration.

\vspace{-1mm}
\begin{tcolorbox}[left=1mm,right=1mm]
    \vspace{-1mm}
    \textbf{Answer to RQ3:} \tool successfully discharges 1,022 out of 1,421 verification targets across diverse real-world verification tasks, outperforming all baselines overall. Its effectiveness spans functional correctness, RTE freeness, and loop invariant generation, demonstrating its ability to synthesize specifications for programs with varying characteristics and verification objectives.
    \vspace{-1mm}
\end{tcolorbox}
\vspace{-1mm}


\subsection{RQ4: User Study}

\begin{table}[]
    \centering
    \caption{Mean, median, and standard deviation of the user ratings obtained by each baseline for both Q1 (Understanding helpfulness) and Q2 (semantic accuracy).} 
    \vspace{-2mm}
    \resizebox{1\linewidth}{!}{
\begin{tabular}{cccccc|cccccc}
\toprule
\multicolumn{6}{c|}{Q1} & \multicolumn{6}{c}{Q2} \\ \midrule
\multicolumn{1}{c|}{Metrics} & AutoSpec & Preguss & SpecGen & \tool & Oracle & \multicolumn{1}{c|}{Metrics} & AutoSpec & Preguss & SpecGen & \tool & Oracle \\ \midrule
\multicolumn{1}{c|}{Mean} & 1.61 & 2.80 & 4.05 & 3.78 & 3.99 & \multicolumn{1}{c|}{Mean} & 1.62 & 2.93 & 3.92 & 4.07 & 9.95 \\
\multicolumn{1}{c|}{Med} & 1.00 & 2.00 & 4.00 & 4.00 & 4.00 & \multicolumn{1}{c|}{Med} & 1.00 & 3.00 & 4.00 & 4.00 & 4.00 \\ 
\multicolumn{1}{c|}{SD} & 1.22 & 1.28 & 1.10 & 1.28 & 1.10 & \multicolumn{1}{c|}{SD} & 1.21 & 1.25 & 1.06 & 0.96 & 1.01 \\ \bottomrule
\end{tabular}
    }
    \label{tab:rating}
\end{table}

We further conduct a user study to investigate how human programmers perceive the specifications' quality. For each of the 14 repositories mentioned in section~\ref{sec:benchmark}, we randomly select one function for which all baselines could generate specifications, and arrange 17 participants (9 M.S. students and 8 Ph.D. students, with at least 3 years of programming experience, all trained on ACSL and Frama-C/WP through relevant coursework) to review all 70 code-specification pairs (14 functions $\times$ 5 baselines, including ground truth). Two questions are prepared for each pair:
\begin{itemize}[leftmargin=*]
    \item Q1: To what extent do the specifications facilitate users in understanding program semantics?
    \item Q2: How effectively do the specifications capture the intended functionalities of the program?
\end{itemize}
The answers are given as 5-point Likert-Scale ratings. To enable fair comparison, the participants are not previously informed of the name or number of the baselines evaluated, and the source baseline for each code-specification pair is anonymized in the questionnaire. 

\begin{figure}
    \centering
    \includegraphics[width=0.85\linewidth]{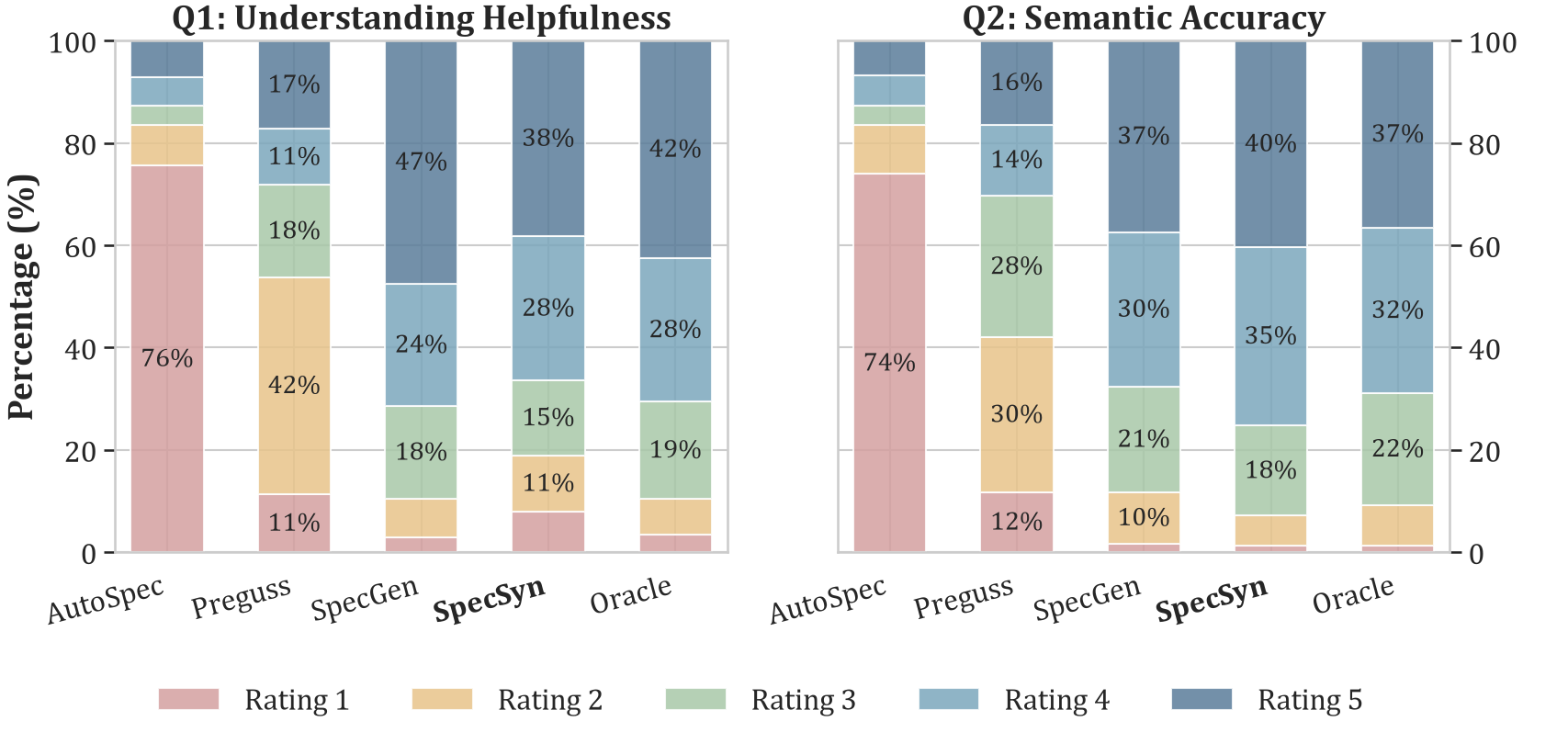}
    \vspace{-3mm}
    \caption{User rating (1$\sim$5) distributions of each baseline method. Percentages below 10\% are omitted for clarity.}
    \vspace{0mm}
    \label{fig:rating}
\end{figure}

Table~\ref{tab:rating} presents the mean, median, and standard deviation for each method on both dimensions. Three clearly separated tiers emerge. AutoSpec (average rating being 1.61 and 1.62, respectively) sits near the floor with the minimum rating. Preguss occupies the middle tier (average rating being 2.80 and 2.93). The top three methods, SpecSyn (3.78 and 4.07 on average), SpecGen (4.05 and 3.92 on average), and Oracle (3.99 and 3.95 on average), cluster tightly within 0.3 points of each other.  Post-hoc Wilcoxon tests with Bonferroni correction indicate that AutoSpec is significantly worse than all other methods ($p<0.01$), Preguss is significantly worse than all three top-tier methods ($p<0.05$), and no pairwise comparison among SpecSyn, SpecGen, and Oracle reaches significance ($p<0.14$). Fig.~\ref{fig:rating} further shows the distribution of raw scores assigned to each method. The patterns reinforce the tiered structure: AutoSpec concentrates at rating 1 (over 70\%), Preguss peaks at ratings 2–3, while the three top-tier methods skew toward ratings 4–5. Among the top tier, \tool and Oracle exhibit nearly identical rating profiles on Q2, with most ratings falling at 4 and 5. Notably, the Q1 distribution of \tool is slightly more spread (1.28 SD) while its Q2 distribution is the tightest among all methods (0.96 SD), suggesting that participants converge more strongly on the judgment that \tool's specifications are semantically accurate.

\vspace{-1mm}
\begin{tcolorbox}[left=1mm,right=1mm]
    \vspace{-1mm}
    \textbf{Answer to RQ4:} The user study shows that \tool produces specifications that are well received by human programmers in both semantic comprehensibility and functional accuracy. Its ratings are comparable to those of SpecGen and the ground-truth specifications, with no statistically significant differences among the three methods. Participants exhibit particularly strong agreement on the functional accuracy of \tool's specifications, as reflected by their low rating variance. These results suggest that \tool generates specifications that are not only formally verifiable but also meaningful and useful for human understanding.
    \vspace{-1mm}
\end{tcolorbox}
\vspace{-1mm}


\section{Discussion}


\subsection{Limitations}

\begin{table}[!t]
    \centering
    \caption{Computational cost (per file) of each baseline.}
    \vspace{-2mm}
    \resizebox{0.6\linewidth}{!}{
    \begin{tabular}{@{}c|cccc@{}}
    \toprule
    Method & Time (s) & LLM Queries & Output Tokens & Verifier Calls \\ \midrule
    \tool & 7068.81 & 147.85 & 42493.36 & 350.20 \\
    Preguss & 5112.08 & 139.19 & 33140.19 & 345.15 \\
    AutoSpec & 594.01 & 4.94 & 2546.81 & 4.94 \\
    SpecGen & 9175.20 & 6.36 & 27626.62 & 587.25 \\ \bottomrule
    \end{tabular}
    }
    \label{tab:overhead}
\end{table}

The main limitation of \tool lies in its computational cost, as shown in Table~\ref{tab:overhead}. Notably, \tool and Preguss involve significantly more LLM queries than other baselines due to the divide-and-conqeur style process in these methods. Compared to Preguss, \tool also consumes more execution time and slightly more verifier calls, since the specification refinement mechanism entails repeated verifier invocations. Nevertheless, the overhead of \tool remains on the same order of magnitude as Preguss. Considering the relatively limited performance of competing approaches (as shown in Table~\ref{tab:RQ1} and~\ref{tab:RQ3}), the general overhead of SpecSyn is acceptable and justified. Also, compared to the substantial human effort required for verification tasks (e.g., 5 calendar months for the manual verification of the X509-parser~\cite{ebalard2019journey}), \tool can substantially alleviate the burden of verifying real-world programs. The limitation can be alleviated by refining the framework implementation, such as further enhancing the prompt design or incorporating concurrent computing. 



\vspace{-1mm}
\subsection{Threats to Validity}

\noindent\textbf{Internal Threats.} One internal threat lies in the potential data leakage of the constructed benchmark. Most programs involved are not susceptible to data leakage, as their ground-truth specifications were manually constructed as part of this work. Only a small subset of test cases inherited from prior work (e.g., those related to the X509-parser project) may be susceptible to data leakage.
Another threat concerns the inherent stochastic nature of LLMs. To mitigate this threat, we adopt greedy decoding (temperature set to 0) whenever possible to ensure replicability, while using the default configuration for models whose parameters are not controllable. Another threat lies in the scale of the constructed benchmark. Only 50 programs are collected due to the substantial manual effort required to construct high-quality ground-truth specifications that satisfy the desired verifiability and strength. To mitigate this threat, we additionally we additionally evaluate the generated specifications on program verification tasks, leveraging existing benchmarks such as SV-COMP to assess their effectiveness across a broader set of programs. We will explore leveraging specification synthesis techniques and coding agents to assist in the construction of larger-scale benchmarks. Finally, our evaluation measures specification recall against manually constructed ground-truth specifications, which may not represent the only valid expression of a target program's semantics. Generated specification may differ syntactically or structurally from the ground truth while still capturing the intended semantics. To mitigate this threat, rather than relying on exact matching, similarity-based metrics, or LLM-as-a-judge, we follow existing works to construct composite specifications based on bidirectional implication and leverage a verifier to establish the semantic equivalence between LLM-generated and ground-truth specifications. This approach reduces the risk of underestimating recall due to differences in specification representations.

\noindent\textbf{External Threats.} The main external threat lies in the verifier adopted for metric calculation, i.e., Frama-C/WP. Despite being a mature verification framework for ACSL, it may be affected by limitations of its implementations and underlying solvers. Some semantically correct specifications may fail to be automatically discharged. Nevertheless, we consistently employ the same tool and configuration for all compared approaches, ensuring that any potential verifier imperfections affect all baselines equally, thereby guaranteeing fairness. Also, our evaluation is limited to C programs annotated with ACSL and verified using Frama-C/WP. The proposed agentic workflow may be influenced by the language-specific features and verification mechanisms of this ecosystem, potentially limiting the generalizability of our findings to other programming languages, specification languages, and verification tools. We plan to explore the application of \tool on other programming languages in future work.

\section{Related Works}

\noindent\textbf{Program Mutation for Testing.} One typical use of program mutations is mutation-based testing, which measures the effectiveness of a test suite by assessing its ability to detect potential faults~\cite{jia2010analysis,papadakis2019mutation}. The operators typically involve only simplistic modifications targeting only a single program point~\cite{coles2016pit,wang2021faster,wang2024exploratory}. Apart from mutation-based testing, program mutations also play a vital role in compiler fuzzing. Conventional compiler fuzzing techniques typically rely on manually crafted mutation operators~\cite{le2014compiler,le2015finding,sun2016finding}. With the rise of LLMs, recent works have proposed diverse fuzzing approaches incorporating LLMs into program mutation~\cite{deng2023large,ou2024mutators,munley2024llm4vv}. In this work, we leverage program mutation to assess and refine specification strength, rather than to perform testing.

\noindent\textbf{LLM-based Specification Generation.}
The widespread application of LLMs has brought about a comprehensive transformation in the field of specification generation. AutoSpec~\cite{wen2024enchanting} first incorporates LLMs into specification generation tasks and utilizes static analysis to identify code hierarchy as guidance to LLMs. SpecGen~\cite{ma2024specgen} introduces conversational generation and heuristic mutation operators in an attempt to repair verification errors of LLM-generated results. Pei et al.~\cite{pei2023can} and Cao et al.~\cite{cao2025informal} aim to improve the performance of LLMs on specification generation tasks by fine-tuning them on data corpora related to program specifications. Compared to these works that treat the input program holistically, \tool handles the input program in a divide-and-conquer manner, enhancing the scalability of the approach. Preguss~\cite{wang2025tale} incorporates abstract interpretation to extract potential runtime errors (RTE), guiding the LLM to generate specifications that exclude these errors. Compared to this work, which focuses solely on RTE-related specifications, \tool adopts a more versatile perspective, independent of the target specification's type, and can generate specifications that characterize the essential behaviors and functionalities of the input program. Other works~\cite{chen2024automated,liu2024propertygpt,yang2025autoverus,zhong2025rag} also target other formal ecosystems, such as Verus specifications for Rust code. Due to intractable differences on target programming language, specification language, and the corresponding verifier, these works are excluded from the comparison. Among these works, some have attempted to evaluate specification strength through program mutations~\cite{amaral2026mutdafny,endres2024can}. Compared to these works that stops on a mere assessment result, we further formulate the counterexample-guided refinement mechanism that utilizes the assessment results to enhance specification strength. To the best of our konwledge, this is the first work to explore ACSL specification strength refinement mechanism for C programs.

\section{Conclusion}


This paper presents \tool, a novel LLM-based specification generation approach. \tool incorporates top-down decomposition and bottom-up synthesis to enhance the framework's scalability, while innovating a specification refinement mechanism based on program mutation and variant discrimination. Experimental results demonstrate the quality of specifications generated by \tool, the effectiveness of the innovated specification refinement mechanism, and the applicability of \tool on real-world verification tasks.


%
%
%
\bibliographystyle{ACM-Reference-Format}
\bibliography{ref}
\newpage

\end{document}